\documentclass[aps,prl,twocolumn,preprintnumbers,showpacs,amsmath,amssymb]{revtex4-1}
\usepackage[colorlinks=true,linkcolor=blue,citecolor=blue,urlcolor=blue]{hyperref}
\usepackage[utf8]{inputenc}
\usepackage{graphicx}
\usepackage{verbatim}
\usepackage{bm,footnote}

\begin{document}
\preprint{APS/123-QED}

\title{Stability of networks of delay-coupled delay oscillators}

\author{Johannes M. H\"ofener}
\email{hoefener@mpipks-dresden.mpg.de}
\author{Gautam C. Sethia}
\altaffiliation{Institute for Plasma Research, Bhat, Gandhinagar 382 428, India}
\author{Thilo Gross}
\affiliation{%
Biological Physics Section, Max-Planck-Institut f\"{u}r Physik komplexer Systeme, N\"{o}thnitzer Stra\ss e 38, Dresden 01187, Germany}%
\date{\today}

\begin{abstract}
Dynamical networks with time delays can pose a considerable challenge for mathematical analysis.
Here, we extend the approach of generalized modeling to investigate the stability of large networks of
delay-coupled delay oscillators. When the local dynamical stability of the network is plotted as a
function of the two delays then a pattern of tongues is revealed. Exploiting a link between structure
and dynamics, we identify conditions under which perturbations of the topology have a strong impact on 
the stability. If these critical regions are avoided the local stability of large random networks 
can be well approximated analytically.
\end{abstract}


\maketitle

Dynamical networks with time-delays (delay networks) have many applications in diverse range of fields from physics and biology \cite{May1973,Roos1991,Eggert2001,DeJong2002,Arenas2008,Englert2010,Atay2010,Callan2010}. 
In particular in biological systems, both the functional forms governing individual dynamical elements and the precise topology of interdependencies are often uncertain. 
For making progress it is therefore crucial to gain a general understanding which properties of the local dynamics and the coupling topology have a strong impact on the system-level dynamics.  

The analysis of delay systems is challenging, because even a single delay-differential equation (DDE) constitutes an infinite-dimensional dynamical system.
In the past, investigations of delay networks in continuous time have primarily focused on small or structurally simple systems \cite{Schuster1989,Dodla2004,Konishi2008,DHuys2010,Shrimali2010}.
Numerical explorations of larger networks emphasized different behavior can be observed depending on the coupling topology \cite{Ponce2009}, which can be related to the spectrum of the graph Laplacian \cite{Atay2004,Atay2006JDE}.
Using a different approach the effect of random delays was analyzed successfully by a mean-field approximation \cite{Gong2008}.

For a variety of dynamical systems without delay it was recently shown that the dynamics of large networks can be analyzed efficiently by the approach of generalized modeling \cite{Gross2006,Gross2009}. 
In particular this approach was recently applied to systems from ecology \cite{Stiefs2008,Yeakel2011} and cell biology \cite{Steuer2006,Zumsande2010}. 
In this context, the ability to incorporate delays into generalized models is highly desirable.   

We consider a class of models where time delays appear both in each network node and in the coupling of nodes.
This is inspired by earlier works that showed that the interplay of two different delays in a single node \cite{Ahlborn2004}, or in the coupling \cite{Konishi2010}, gives rise to rich dynamics.  
Both intra-node and coupling delays are present in ecological metacommunities (times needed for maturation of individuals and migration between patches) and systems biology (protein assembly time, active/passive transport times).

Here we use generalized modeling to obtain an expression governing the stability of all stationary states in this class of system.
A numerical analysis reveals a rich pattern of tongues of different instabilities in the space spanned by the two delays.
We then propose an approximation, allowing for the analytical investigation of the pattern of instability. 
Thereby, we identify the conditions under which small variations in the network topology have a strong impact on stability.  
If these regions are avoided then the local stability of large networks of delay-coupled delay oscillators can be well approximated analytically.

\section{Delay-coupled delay networks\label{secModel}}
We consider networks of $N$ nodes and $K$ bidirectional links. 
The topology of these networks is captured by the adjacency matrix $\mathrm{\mathbf{A}}$, such that $A_{i j}$ is 1 if nodes $i$ and $j$ are connected and 0 otherwise.
Each node $i$ has an internal dynamical variable $X_i$, representing for instance the abundance of an ecological population or the density of mRNA molecules.
Hence, $X_i$ changes due to internal dynamics in $i$ and coupling to the neighbor's variables $X_j$ according to
\begin{equation}\label{eq:DDE}
\dot{X}_i=G(X_i^{\tau})-L(X_i)+\sum_j A_{i j} (F(X_j^{\delta})-F(X_i)),
\end{equation}
where $\tau$ and $\delta$ denote internal and travel-time delays and $G$, $L$ and $F$ are positive functions describing gain, loss, and coupling, respectively. 
In our analytical treatment we do not restrict these functions to specific functional forms, but consider formally the whole class of models, which includes several well-studied examples such as the Mackey-Glass \cite{MackeyGlass1988} and the Ikeda model \cite{Ikeda1987}.

\section{General stability analysis\label{secGeneral}}
The challenge that we address in the following is to determine the local stability of an arbitrary (positive) steady state ${X_i}^*$.
In the physical system such states may either correspond to stationary points (e.g.~oscillation death) or phase-locked oscillations described in a co-moving frame.

Previous works on generalized models showed that it is possible to express the stability of arbitrary steady states as functions of easily interpretable parameters if a certain normalization is used \cite{Gross2006,Gross2009}.  
We therefore introduce normalized variables, $x_i=X_i/X^*$, and normalized functions, $f(x_i)=F(x_i X^*)/F(X^*)$, denoted in lower case.
Using $X^*=X^{\tau *}=X^{\delta *}$, we write
\begin{equation}\label{eq:DDE_norm}
\dot{x}_i=\alpha(g(x_i^{\tau})-l(x_i))+\beta\sum_j A_{i j} (f(x_j^{\delta})-f(x_i)),
\end{equation}
where $\alpha=G(X^*)/X^*=L(X^*)/X^*$ and $\beta=F(X^*)/X^*$ can be interpreted as parameters describing characteristic per-capita rates of growth and transport. 
Finally, we set $\alpha=1$ by normalization of the time scale.

The stability of the normalized steady state $\mathbf{x}^*=\mathbf{1}$ can be determined from a local linearization given by the  Jacobian matrix $\mathrm{\mathbf{J}}$ with $J_{ij}= \partial \dot{X}_i / \partial X_j$.
Close to the steady state, small perturbations can be decomposed into eigenvectors $v_l$ of the Jacobian \cite{Guckenheimer1983}.
The time evolution of a perturbation $y$ along the eigenvector $v_l$, then follows a locally exponential trajectory 
such that   
\begin{equation}
\frac{\partial y^\tau}{\partial y} = {\rm e}^{-\lambda_{l} \tau},
\end{equation}
where $\lambda_l$ is an eigenvalue of the Jacobian, corresponding to the eigenvector $v_l$.
Using this relation we capture the response to a given perturbation by a Jacobian with
\begin{equation}
\label{eq:Jac}
\begin{array}{rclll}
\begin{split}
J_{ii} & =(g' {\rm e}^{-\lambda \tau}-l')-d_i\beta f' & =: & J_i^\mathrm{d},\\
J_{ij} & =\beta f' {\rm e}^{-\lambda \delta} A_{i j}  & =: & J^\mathrm{o} A_{i j},
\end{split}
\end{array}
\end{equation}
where dashes denote derivatives with respect to the argument in $x^*$ and $\lambda$ is a self-consistent eigenvalue of ${\rm \bf J(\lambda)}$. 

The steady state is stable if all of these eigenvalues have negative real parts.  
Conversely, the steady state is unstable if at least one such eigenvalue has positive real part, such that there is a perturbation that grows in time.

We emphasize that Eq.~\eqref{eq:Jac} states the Jacobian of all steady states in all models of the form of Eq.~\eqref{eq:DDE}.
The Jacobian is written as a function of scalar quantities that can be interpreted as unknown parameters. 
Specifically, the parameters $g'$, $l'$, and $f'$ are logarithmic derivatives of the original functions (e.g.~$f'=\left.\partial f/\partial x\right|_1=\left.\partial {\log(F)}/\partial{\rm log}(X)\right|_{*}$), which are known as elasticities and are used in many fields because they often have an intuitive interpretation in the context of the application \cite{Fell1992}.

Because the present analysis focuses primarily on the effect of topology, we restrict ourselves to the case $g'=-1$, $l'=0$, $f'=1$, $\beta=1$. A detailed analysis of the effect of parameters will be published separately.

\begin{figure}[!t]
\includegraphics[scale=2.0]{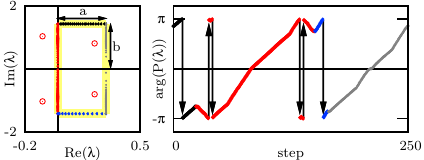}
\caption{(Color online) Computation of the number of eigenvalues with positive real part. Left: Location of the largest eigenvalues of an example Jacobian in the complex plane (circles). The algorithm follows a rectangular contour C (yellow), taking discrete steps (dots). The size of C, marked $a,b$, is chosen sufficiently large that all positive eigenvalues must lie within contour (Gershgorin's theorem). Right: The number of positive eigenvalues is found as the winding number of $\arg P(\lambda)$ as $C$ is followed in positive (counter clockwise) direction. Colors (black/red/blue/gray) make different segments of $C$.\label{Fig:Cauchy}}
\end{figure}
In DDEs the computation of eigenvalues is complicated by the explicit appearance of the eigenvalue $\lambda$ in $\rm \bf J$,
which turns the characteristic polynomial $P(\lambda)$ into an implicit transcendental equation. 
We therefore follow the approach of \cite{Luzyanina1996} and test for eigenvalues with positive real parts (EVPs) using Cauchy's argument principle: 
For analytic functions $P(\lambda)$, the number of roots inside a contour C is $N_{\rm C}=\frac{1}{2\pi}\Delta_C \arg P(\lambda)$ 
where $\Delta_C \arg P(\lambda)$ is the winding number of $P(\lambda)$ on C (Fig.~\ref{Fig:Cauchy}). 
The total number of EVPs can therefore be computed by applying the Cauchy principle to a contour in the positive half-plane that is chosen so large that all EVPs must lie within the contour. 
In the present Letter we use a rectangular contour covering the interval $[0,a]$ in the real and $[-b,b]$ in the imaginary direction, where $a$ and $b$ are estimated using Gershgorin's theorem \footnote{All eigenvalues lie within a set of $N$ circles, where circle $i$ is centered on $J_{ii}(\lambda)$ and has the radius $r_i=\sum_{j\ne i} |J_{ij}(\lambda)|$. The center itself lies on a circle of radius $\rho_i=|g'|e^\kappa\tau$ around $M_i=-l'-d_i\beta f'$, where $\kappa=\mathrm{Re}(\lambda)$. We find an upper bound for $a$ by numerically solving the implicit equations $M_i+R_i(a_i)=a_i$, with $R_i=r_i+\rho_i$. An upper bound $b$ is given by $b=\max_i\left(\sqrt{R(0)^2-M_i^2}\right)$.}.
\begin{figure*}[!ht]
\includegraphics[]{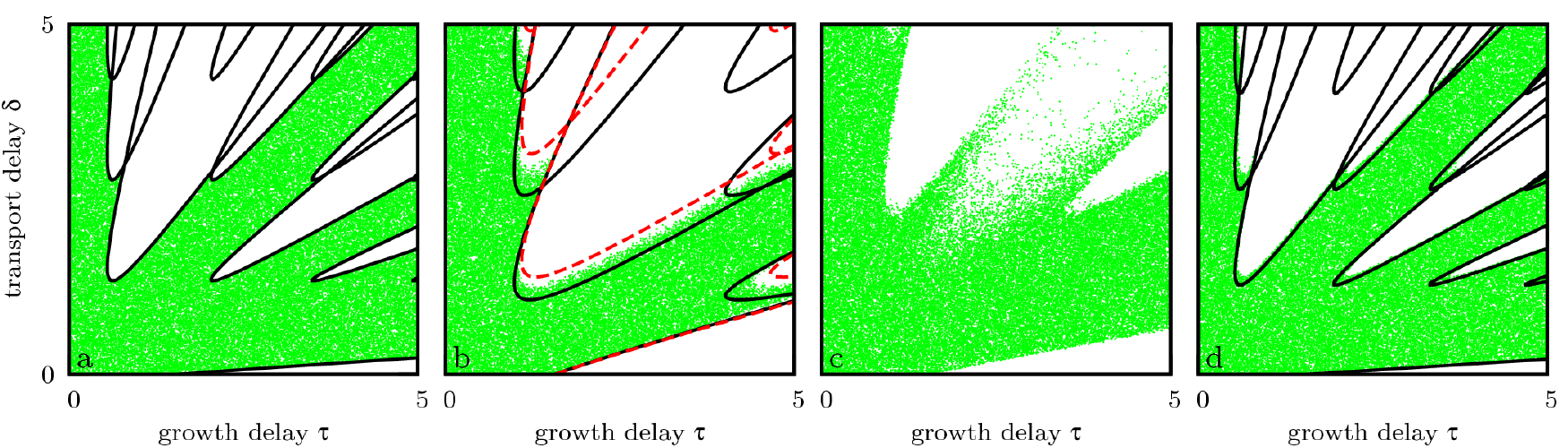}
\caption{(Color online) Local stability and bifurcations of networks. The parameter space in the panels is sampled uniformly. Stable states are marked in green, whereas unstable states are not shown. Panels show stability in the fully-connected network with $N=10$ nodes and $K=45$ links (a), random trees with $N=10$ and $K=9$(b), and random networks $N=10$ and $K=15$ (c), and $N=100$ and $K=500$ (d). Areas of instability (white) are sharply delineated if a single topology is considered (a). In network ensembles fuzzy regions appear, where stability is topology dependent (b,c). In case of random trees these regions are bounded by the bifurcation lines of star (dashed) and chain (solid) topologies (c). However, even in large random networks with many links the regions of instability have relatively sharp boundaries, which can be approximated by analytical bifurcation lines for the fully-connected network of identical degree $d=10$ (d, solid).}\label{Fig:Tongues}
\end{figure*}
For efficient computation of the winding number we use an adaptive step-size algorithm counting the crossings of odd multiples of $\pi$. 
We iterate along C by stepping from a point $\lambda_i$ to the subsequent point $\lambda_{i+1}=\lambda_i+h v$, where $v$ is a unit vector along C and $h$ is the current step size. 
The step is accepted if a) $\Delta:=D(\arg P(z_i)-\arg P(z_{i+1})))<\epsilon=0.1$ and b) $D(\arg P(z_i)-\arg P((z_i+z_{i+1})/2))<\Delta$ where $D(x)=|x| \bmod{2\pi}$. 
Otherwise, the step is rejected and $h\to h/2$. 
After each successful step $h \to h \max(2,\epsilon/\Delta)$.
 

\section{Stability in fully-connected networks \label{secFullNet}}
We now explore the effect of topology and delay on local stability of steady states. 
For understanding the effect of delays on dynamical stability we first consider a fully-connected network. 
The stability of this network is explored numerically by generating an ensemble of parameter sets, where the delays $\tau$ and $\delta$ are drawn randomly from a uniform distribution.
The stability of the steady states corresponding to the sample parameter sets is then evaluated by the method described above. 
Because of the efficiency of this method the evaluation of the ensemble only requires minutes of computational time. 

The result of the sampling analysis reveals a pattern of tongues of instability,
which are separated by stable channels located around the resonant delays $\tau=n \delta$ with integer $n$ (Fig.~\ref{Fig:Tongues}a).

We note that these results differ from those found in systems with two delays of the same type \cite{Ahlborn2004, Konishi2010}, where resonance takes the form $m \tau = n \delta$.

Qualitative changes in the dynamics, including changes in the stability can only occur when the system undergoes a bifurcation.  
For the fully-connected network we computed the local bifurcation points explicitly by numerical continuation of the bifurcation condition $\mathrm{Re}(\lambda)=0$. 
Some of the bifurcations that are thus revealed mark changes of stability on the edge of the tongue, whereas others correspond to qualitative transitions within the unstable region.  

\begin{figure}[!t]
\includegraphics[scale=2.0]{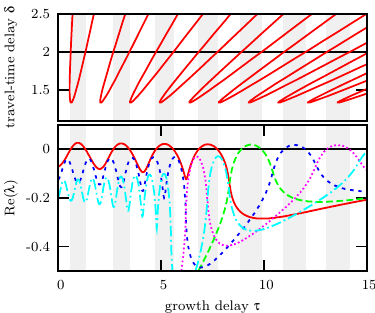}
\caption{(Color online) Lattice symmetry in the bifurcation points. Top: Blowup of a part of the bifurcation diagram, Fig.~\ref{Fig:Tongues}a. Bottom: Selected eigenvalues of the Jacobian (different colors) along a one dimensional cut at $\delta=2$ (black line, top). Shaded regions mark values of $\tau$ where the system with $\delta=2$ is unstable.}
\label{Fig:EvReal}
\end{figure}

Considering the tongues in Fig.~\ref{Fig:Tongues}a further, one notices that the tips of the tongues are located on the vertices of a square lattice.
A similar symmetry was already observed previously in simpler models \cite{Yanchuk2009} and can be explained as follows: 
Because the edge of a tongue is a local bifurcation, the Jacobian has to have a purely imaginary eigenvalue $\lambda=i\omega$.   
Moreover, a delay parameter, say $\tau$, can only appear in the Jacobian in factors of the form ${\rm e}^{-\lambda\tau}$. 
Therefore, increasing the delay by a multiple of $2\pi/\omega$ leaves the Jacobian invariant because ${\rm e}^{-i\omega (\tau + 2 \pi / \omega)}={\rm e}^{-i\omega \tau}$, i.e., when starting from a bifurcation point ($\tau,\delta,\omega$), increasing either of the two delays by $2 \pi/\omega$ must lead to another bifurcation point. 
This proves that not only the tips of the tongues, but every bifurcation point is part of a square lattice. 
For understanding why the periodicity is most visible in the tips, consider that different points on the edge of a single tongue generally differ in $\omega$ and therefore also in the corresponding lattice constant.
Therefore, the tongues as such do not reappear identically at higher delays but become distorted. 
We emphasize that this symmetry cannot be extended to points in parameter space that are not bifurcation points (see Fig.~\ref{Fig:EvReal}).

\section{Stability in random trees}
We investigate the impact of network topology on the stability first by considering an ensemble of random tree topologies.
For any given tree the analysis described above reveals sharply delineated tongues of stability.
The number and location of these tongues depends on the specific topology under consideration.
For visualizing the differences and similarities between tongue patterns, we repeat the sampling analysis described above but also draw a random tree topology for every individual sample point.     
Because results from many different topologies are thus superimposed in Fig.~\ref{Fig:Tongues}b, the edges of the tongues become fuzzy. 
The fuzzy areas thus mark parameter regions where the stability depends on the network topology.
Nevertheless, large region exist in which the stability remains independent of topology. 
 
We found empirically that the stability boundaries for all tree topologies fall between those for the linear chain and the star topology, which are the least and most dynamically stable topologies, respectively. 
In some places the stability boundaries of even these extreme cases coincide so that the corresponding transitions occur independently of the specific tree topology.

In contrast to the fully-connected network, the periodicity in the travel-time delay, $\delta$, is reduced to $\pi/{\omega}$ for trees. The resulting additional tongues cover the stable channel around $\tau=\delta$, present in the fully-connected network. 

\section{Stability in random networks}
For further exploration of the effect of topology and understanding the appearance of additional tongues of instability in the random trees, we now consider an ensemble of Erd\"os-Reny\'{i} random graphs, with fixed number of links, $L$.

Below, we distinguish two $2\pi$-periodic sets of tongues: the off-diagonal set (OS), which is already present in the fully-connected network, and the diagonal set (DS), which appeared in the trees.
By visual inspection of the phase diagrams for different topologies we observed that the DS is present in some networks, but is absent in others. 

While the $2\pi$ periodicity within the sets is guaranteed by analytical arguments, the relative offset between the sets can depend on topology. 
The $\pi$-periodicity, observed in the random trees, requires that the DS is shifted relative to the OS exactly by $(0,\pi/\omega)$ in $(\tau,\delta)$.
We find this particular offset whenever the network topology is bipartite, i.e.~when the network can be colored with two colors such that no link connects nodes of the same color. 
The observed $\pi$-periodicity therefore arises because all trees meet the condition of bipartiteness. 

The dependence of the stability on the topology noted earlier has to stem from differences either in the number or location of the tongues of stability. 
Given two network topologies, parameter regions with different stability properties appear because a) the tongues found for one of the topologies are shifted with respect to the tongues found for the other topology, or b) tongues of the DS are present in one of the topologies but absent in the other.  
In comparison, the effect of a) is relatively minor, causing for instance the small regions of topology dependence in Fig.~\ref{Fig:Tongues}b.
By contrast, if $K$ is tuned to a value where the DS appears, stability can be topology dependent in large regions of the parameter space (Fig.~\ref{Fig:Tongues}c).
 
Even when ensembles of large random networks are considered, we observe that the tongues of instability remain relatively sharply delineated (Fig.~\ref{Fig:Tongues}d) unless parameters are tuned to regions of parameter space where new tongues appear (Fig.~\ref{Fig:Tongues}c).  
This implies that in the ensemble considered here, the dynamical stability is to a large extend independent of the specific network topology. We note that the stability of large and dense random networks closely matches the results for a fully-connected network with the same mean degree (Fig.~\ref{Fig:Tongues}d). For the special case of degree-homogeneous networks, this observation is explained below.


\section{Analytical theory}
For gaining an analytical understanding of the results presented above, we consider the case of degree-homogeneous networks, in which all nodes have the same number of connections. 
For these networks the eigenvalues of the Jacobian from Eq.~\eqref{eq:Jac} are given by the implicit equation
\begin{equation}\label{eq:ev_approx}
\lambda=J^\mathrm{d}(\lambda)+c J^\mathrm{o}(\lambda),
\end{equation}
where $c$ can be any eigenvalue of the adjacency matrix, which we denote as \emph{topological eigenvalues}.
Bifurcations are characterized by the presence of a purely imaginary eigenvalue $i \omega$. 
This leads to
\begin{equation}
\label{eq:ReIm}
\begin{split}
0&=g'\cos(\phi)-l'-d\beta f'+c\beta f'\cos(\psi),\\
\omega&=-g'\sin(\phi)-c\beta f'\sin(\psi).
\end{split}
\end{equation}
where we used Eq.~\eqref{eq:Jac} and $\phi:=\omega\tau$, $\psi:=\omega\delta$.

For analyzing the bifurcation condition we enumerate its solution branches. 
First, we note that for each solution triplet $(\phi,\psi,\omega)$, there exist another solution $(-\phi,-\psi,-\omega)$ corresponding to the identical tongue. 
Therefore, we only consider solutions with $\omega\ge0$. 
Second, given a solution $(\phi,\psi,\omega)$ other solutions are found at $(\phi+2\pi r, \psi+2\pi s,\omega)$, where $r,s$ are integers enumerating the tongues. 
Finally, if there is a triplet $(\phi,\psi,\omega)$ that solves Eq.~\eqref{eq:ReIm} for $c<0$, there must be a triplet $(\phi,\psi+\pi,\omega)$ that solves modified equations in which $c$ is replaced by $|c|$. 
We can therefore enumerate the solutions with negative $c$ by half-integer values of $s$ in the modified system.

Considering the first tongue ($r=s=0$) we find
\begin{equation}
\label{eq:psiomega}
\begin{split}
\psi&=2\pi\pm\cos^{-1}(p(\phi))\\
\omega&=-g'\sin(\phi)\mp |c|\beta f' \sqrt{1-p(\phi)^2},
\end{split}
\end{equation}
where
\begin{equation}
\label{eq:p}p(\phi)=\frac{d}{|c|}+\frac{l'-g' \cos(\phi)}{|c|\beta f'}.
\end{equation}
and the codomain of $\cos^{-1}$ is $[0,\pi]$. 
These equations provide a representation of the tongues depending on $\phi$.

We define the tip of a tongue as the bifurcation point recurring with the smallest lattice constant. 
The tips are thus characterized by maximal values of $\omega$, such that
\begin{equation}
\label{eq:tip}
\begin{split}
\tau_\mathrm{tip}&=\frac{\cos^{-1}(q)+2r\pi}{(|c|\beta f'-g')\sqrt{1-q^2}},\\
\delta_\mathrm{tip}&=\frac{\cos^{-1}(q)+(2s+1)\pi}{(|c|\beta f'-g')\sqrt{1-q^2}},
\end{split}
\end{equation}
with $q=-(d\beta  f'+l')/(|c|\beta f'-g').$ Eq.~\eqref{eq:tip} shows that half-integer values of $s$, or equivalently topological eigenvalues $c<0$, correspond to the DS, while integer values of $s$ correspond to the OS.
Further, Eq.~\eqref{eq:tip} requires $|q|<1$ and hence
\begin{equation}
\label{eq:c_cond}|c|>d+\frac{g'+l'}{\beta f'}.
\end{equation}
Topological eigenvalues $c$ violating this condition cannot satisfy the bifurcation condition and hence do not correspond to tongues of instability. 
The positive $c$ creating the OS, and the negative $c$ creating the DS are therefore separated by a forbidden region in which topological eigenvalues do not create tongues. If a change of parameters causes a $c$ to enter this region, the corresponding set of tongues vanishes as the corresponding bifurcation lines shift to infinite delays. 

For understanding the appearance of $\pi$-periodicity in $\delta$ observed in bipartite networks, consider that all bipartite networks have symmetric topological spectra \cite{Cvetkovic1980}. 
Therefore, for every $c>0$ there is a symmetric $c<0$, such that the $\pi$-periodicity appears according to Eq.~\eqref{eq:ReIm}.

Above we observed that the stability of large and dense random networks closely matches the results for the fully-connected network with the same mean degree. This dependence of the similarity on the mean degree can be explained in degree-homogeneous networks as the largest eigenvalue in these networks is $c_{\rm max}=d$. The OSs created by the leading eigenvalues of two different degree homogeneous networks therefore match if the networks have the same degree. Further, the observed similarity requires that no other sets of tongues are present, which seems to be generally true for sufficiently large and dense random graphs and fully-connected networks. This explains that fully-connected networks with appropriately chosen mean degree offer a good approximation for the stability in a large class of random graphs with sufficiently narrow degree distribution.

\section{Numerical simulations}
For confirming the results of the generalized model and illustration of the dynamical implications of bifurcation lines 
we simulate 10 Mackey-Glass systems coupled in a ring topology. 
The simulations use the equations
\begin{equation}
\dot{X}_i=\frac{aX_i^\tau}{1+(X_i^\tau)^b}-c X_i+\epsilon (X_{i-1}^\delta+X_{i+1}^\delta-2 X_i),
\end{equation}
with $a=2,b=10,c=1,\epsilon=10$, corresponding to the general parameters $g'=-4,l'=1,f'=1,\beta=10$.
\begin{figure}[!t]
\includegraphics[scale=2]{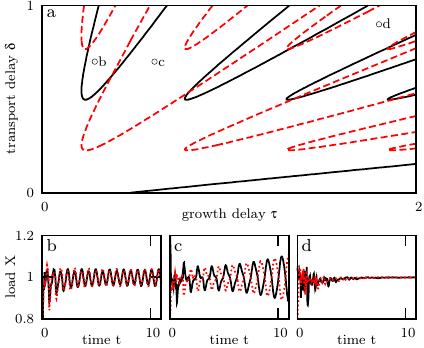}
\caption{(Color online) Comparison between simulations and stability analysis for a ring of 10 Mackey-Glass systems. a) Bifurcation lines of the topological eigenvalues $2$ (solid black) and $-2$ (dotted red). (b-d) Simulation results for two neighboring nodes with $\tau=0.28,\delta=0.6$ (b), $\tau=0.6,\delta=0.6$ (c) and $\tau=1.8,\delta=0.9$ (d). Delay values inside tongues of the eigenvalue $2$ give rise to in-phase oscillations, values inside tongues of eigenvalue $-2$ give rise to anti-phase oscillations. Simulations are performed with the pydelay-package \cite{Flunkert2009}.}
\label{Fig:Sim}
\end{figure}
The two topological eigenvalues $c_1=2$ and $c_{10}=-2$ satisfying Eq.~\ref{eq:c_cond} give rise to one DS and one OS of tongues. 
The dynamical implications of the tongues is illustrated in Fig.~\ref{Fig:Sim}. 
Choosing parameters inside the OS results in in-phase synchronized oscillations, parameters inside the DS, result in anti-phase oscillations, while parameter values outside the tongues result in stationary dynamics. 

\section{Discussion}
In this paper we studied the local dynamics of a large class of networks of delay-coupled delay differential equations. 
For these we showed that the stability of steady states is governed by two distinct sets of tongues of instability.
Further, exploiting a link between network structure and dynamics, we derived an analytical expression for the tongues in degree-homogeneous nets. 

Our results were obtained using the approach of generalized modeling that constitutes a local analysis. 
Presently extensions of this approach to nonlocal dynamics are being developed \cite{Kuehn2011}.
However, even the present local analysis can reveal some insights in nonlocal phenomena. 
For instance it is known that chaotic dynamics are generically present close to double Hopf bifurcations that are found at the 
intersection of tongues \cite{Kuznetsov1995, Gross2005}. 

The main result of our exploration is that the stability in large random networks is relatively independent of the network topology. 
This result seems to be in conflict with previous analysis of other systems which suggest that topological properties such as clustering and cliques can have a strong impact on the dynamics \cite{Atay2006JDE,Ponce2009}. 
The apparent disagreement is resolved if one considers that such topological properties are exceedingly rare in the ensemble of random networks considered here. 

Beyond the numerical results for random networks, our analytical calculations showed that the dynamics are in general only strongly dependent on topology if a variation of the topology brings new tongues of instability into existence. 
This is the case if the variation shifts eigenvalues of the adjacency matrix into a certain range. 
This observation is consistent with previous results as topological properties that have been implicated as influencing the dynamics are closely linked to spectral properties of the networks \cite{Goh2001, Farkas2001, Dorogovtsev2003, Kamp2005}.
We can therefore conjecture that the strong topology dependence of the dynamics observed previously is linked to specific eigenvalues of the adjacency matrix that appear for instance in networks containing many clusters or cliques.
It was recently shown that the appearance of these eigenvalues is not linked to the clusters or cliques themselves, but rather to specific local symmetries that typically accompany them \cite{MacArthur2009}. 
  
If true, the conjecture stated above opens an intriguing possibility: At least in the class of networks considered here, the local dynamics of very different networks with the identical mean degree may indeed be very similar except for additional instabilities which are caused by local symmetries within the networks.

\end{document}